\documentclass[preprint,onecolumn,epsfig,showpacs,preprintnumbers,amsmath,amssymb]{revtex4}
\usepackage{graphicx}% Include figure files
\usepackage{slashbox}

\begin{document}

\title{Numerical study of two-body correlation in a 1D lattice with perfect blockade}

\author{B. Sun and F. Robicheaux}

\affiliation{Department of Physics, Auburn University, Auburn, AL
36849, USA}

\begin{abstract}
We compute the dynamics of excitation and two-body correlation for
two-level ``pseudoatoms" in a 1D lattice. We adopt a simplified
model where pair excitation within a finite range is perfectly
blocked. Each superatom is initially in the ground state, and then
subjected to an external driving laser with Rabi frequency
satisfying a Poissonian distribution, mimicking the scenario as in
Rydberg gases. We find that two-body quantum correlation drops
very fast with the distance between pseudoatoms. However, the
total correlation decays slowly even at large distance. Our
results may be useful to the understanding of Rydberg gases in the
strong blockade regime.
\end{abstract}

\pacs{03.67.Mn, 32.80.Rm}

\maketitle

Recently there have been many experimental efforts in
investigating Rydberg gases
\cite{gallagher,pillet,gould,raithel,pillet2,pfau,Heuvell},
spurred largely by fast developments in laser cooling and
trapping. In such systems, the dipolar interaction between two
nearby atoms will shift the pair excitation out of resonance with
the driving laser. Local excitation is then greatly suppressed,
showing the well-known phenomenon of dipole blockade. It is an
important mechanism responsible for the explanation of many-body
effects such as spectral line broadening \cite{gallagher,pillet}
and the expected sub-Poissonian atom counting statistics
\cite{raithel,rost,francis}. It is also proposed to utilize dipole
blockade to implement quantum gates useful in quantum computation
and information \cite{lukin,zoller}. To date, there have been many
theoretical proposals and numerical simulations in exploring the
excitation dynamics and coherence properties such as pair
correlation and number correlation, which demonstrate the
important role that dipole blockade played in these systems.

Previous investigations of Rydberg gas have focused on the mean
number of excitations and how the excitations are correlated in a
gas. However, there hasn't been a study of genuine quantum
correlation. This paper will discuss other useful information that
can be extracted from the wave function, especially the two-body
quantum correlation and total correlation (including classical
correlation). As we will see later, two-body correlation is
another important quantity for characterizing such systems. To
investigate correlation properties, we are mainly interested in
the following two questions. First, in such systems, how does
correlation behave as the excitation begins to build up? Second,
which correlation is more important? It has been shown that a mean
field approach is not appropriate in dealing with correlation and
a full quantum calculation is necessary \cite{Robicheaux}.
However, the direct simulation of correlation is still absent due
to numerical complexities in real Rydberg gases.

We first consider the scenario as in current experiments where
Rydberg atoms are confined in a volume $\mathcal{V}$ and driven by
a laser. In the spirit of the dipole blockade model, we can divide
$\mathcal{V}$ into equally sized regions $\mathcal{V}=\bigcup_k
v_k$ with each region approximated as a sphere of blockade radius
$R_b$. We follow the usual notation and call the ensemble of atoms
in each region $v_k$ a ``superatom" \cite{vuletic}. For van der
Walls interaction and a narrow band laser, $R_b$ is determined
from $C_6/R_b^6=\Omega$, where $C_6$ is the van der Waals
coefficient in the atomic unit and $\Omega$ is the Rabi frequency
of the driving laser. Rydberg atoms are randomly distributed in
$\mathcal{V}$ so that the number of atoms in each $v_k$ is in
general different. For example, in the homogeneous case, the atom
numbers in each superatom follow a Poissonian distribution with
the same mean atom number. Such a picture is useful in explaining
certain experimental results, e.g., the population dynamics under
a driving laser. In this case, each superatom evolves with
different Rabi frequencies. This leads to a fast and almost linear
increase in the initial stage and saturation in the long time.
However, such a model is not so interesting in showing the
two-body correlation properties because correlation between
different superatoms is apparently 0. Therefore, a proper
modification to the superatom model is necessary in order to give
nontrivial information on two-body correlation.

\begin{figure}[htb]
\includegraphics[width=1.45in,angle=270]{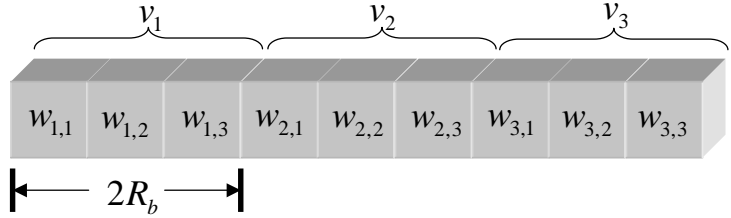}
\caption{A schematic view of partitioning 3 superatoms into 9
pseudoatoms. Each $v_k (k=1,2,3)$ represents a superatom which is
further divided into 3 pseudoatoms $w_{k,\alpha} (\alpha=1,2,3)$.}
\label{scheme}
\end{figure}

In this paper, we discuss an elongated Rydberg gas whose
transverse size is smaller than $R_b$. This allows us to treat the
Rydberg gas as a quasi 1D system. The advantage is two fold. On
the one hand, the edge effect in numerical simulation is smaller
and the results converge relatively faster than those in higher
dimensions with the same atom number. On the other hand, it
permits an easier readout of the quantum state for ensemble of
atoms in any region along the longitudinal direction. Similar to
the 3D case, we divide such a quasi 1D gas into a collection of
superatoms aligned along the longitudinal direction. We
investigate the two-body correlation by further dividing each
superatom ($v_k$) into smaller subregions labelled as
$w_{k,\alpha}$, i.e., $v_k=\bigcup_{\alpha=1}^{N_w} w_{k,\alpha}$.
The center of $w_{k,\alpha}$ is denoted as $r_{k,\alpha}$. The
number of partitions in each $v_k$, labelled as $N_w$, is assumed
the same, $\forall k$. For the interest of this paper, we term
each subregion $w_{k,\alpha}$ as a ``pseudoatom" located at
position $r_{k,\alpha}$. in Fig. \ref{scheme}, we give a schematic
view of partitioning 3 superatoms into 9 pseudoatoms. The
subscripts $k$ and $\alpha$ can be further combined into a single
one by relabelling them as in a spin chain. Interaction between
two pseudoatoms on site $j$ and $k$ are assumed to be a van der
Waals interaction with the distance $|r_j-r_k|$. This
approximation neglects the position variance of excitation in each
pseudoatom. By partitioning each superatom into more pseudoatoms,
the error can be reduced. Different from the case of nearest
neighboring superatoms (with a distance $2R_b$) where the
interaction is usually ignored, we can see that nearest
neighboring pseudoatoms (with a distance of a fraction of $R_b$)
interact much more strongly so all of the pseudoatoms are
correlated analogous to a spin chain.

Due to the $R^{-6}$ dependence of interaction, the transition from
blockade to no blockade has a narrow width of $R$. This gives us
the motivation to assume a perfect blockade with pseudoatoms
within a certain distance and completely no blockade outside of
it. From the many body point of view, the introduction of this
approximation allows us to concentrate on the much simpler short
range interaction rather than the long range interaction. It not
only lets us see the effect due to pure blockade, but also greatly
reduces the needed basis which makes the numerical simulation in a
much larger system possible. The validity of this approximation
will be discussed in the later part of this paper. Because of this
type of partition for a superatom, the maximal number of
pseudoatoms that can be blocked is just the maximal distance
between two pseudoatoms $j$ and $k$ which have an interaction
$V_{j,k}$ much larger than the Rabi frequency $\Omega$, i.e.,
$\max_{j,k}\{j-k|V_{j,k}/\Omega>\eta\}$. Besides the vagueness of
this definition, in practice we choose $\eta=6$. A little algebra
shows that the maximal number of pseudoatoms that can be blocked
is given by $\max_{k=0,1,2...}\{k|k<0.37N_w\}$. The first few
values are given here: for $N_w=2,3,4,5,6,...$, the maximal number
of pseudoatoms that can be blocked is found to be $0,1,1,1,2,...$,
respectively. So the simplest nontrivial situation starts from
$N_w=3$, perfect blockade only between two nearest neighboring
pseudoatoms. In this paper, we will limit our discussion to the
situation where perfect blockade only exists between two nearest
neighboring pseudoatoms.

Because of our approximation, the Hamiltonian can be written as
$\hat{H} = \sum_{k=0}^{N-1} J_k \hat{P}^{(k)}\hat{\sigma}_x^{(k)}
\hat{P}^{(k)}$, with the projector
\begin{eqnarray}
\hat{P}^{(k)} \equiv \prod_{q=k\pm 1} {1-\hat{\sigma}_z^{(q)}
\over 2},
\end{eqnarray}
where $N$ is the total number of pseudoatoms. $\hat{\vec
\sigma}^{(k)}$ are Pauli matrices for the $k$-th pseudoatom.
$J_k=\sqrt{X_k/\lambda}$ is the scaled Rabi frequency for the
$k$-th pseudoatom. We are assuming a resonant laser field with a
constant intensity. $X_k$ is a random variable satisfying a
Poissonian distribution with mean value $\lambda$. In this case,
$\lambda$ denotes the average number of Rydberg atoms in a
pseudoatom. More explicitly, $\lambda = N_r/N_w$ with $N_r$ the
number of Rydberg atoms in a superatom which can be determined
from the Rydberg gas density $n$, the blockade radius $R_b$ and
the area in the transverse direction $A$ through the relation
$N_r=2nR_b A$. The underlying motivation for the assumed form of
$J_k$ is the number fluctuation and the collective excitation in
each pseudoatom \cite{pfau}. A given set of $\{J_0,...,J_{N-1}\}$
is said to define a configuration. Our numerical results are
obtained by averaging over many configurations. We note that the
Hamiltonian formally describes a nonphysical multi-body
interaction, which is of course an effective Hamiltonian
constrained by the perfect blockade requirement. All the
pseudoatoms are initially in the ground state $|g\rangle^{\otimes
N}$, and then subjected to a resonant external laser field. The
subsequent dynamics constrained by the perfect blockade are what
we are after in this paper.

For nearest neighboring blockade, the maximally possible
excitation for all pseudoatoms is found to be $[N/2]$ where $[
\cdot ]$ denotes the integer part. As a consequence, the number of
restricted basis, $N_b$, is significantly reduced from that of the
full basis set. To compute the wave function, we expand it in the
restricted basis labelled as $|\mu_p\rangle$, i.e.,
\begin{eqnarray}
\psi(t) = \sum_{p=1}^{N_b} c_p(t) |\mu_p\rangle,
\end{eqnarray}
the fraction of excitation is found to be
\begin{eqnarray}
P_{ex}(t) = {1\over N} \sum_{p=1}^{N_b} |c_p(t)|^2  \langle
\mu_p|\sum_{k=0}^{N-1} |e\rangle_k \langle e| |\mu_p\rangle.
\end{eqnarray}

\begin{figure}[htb]
\includegraphics[width=3.15in]{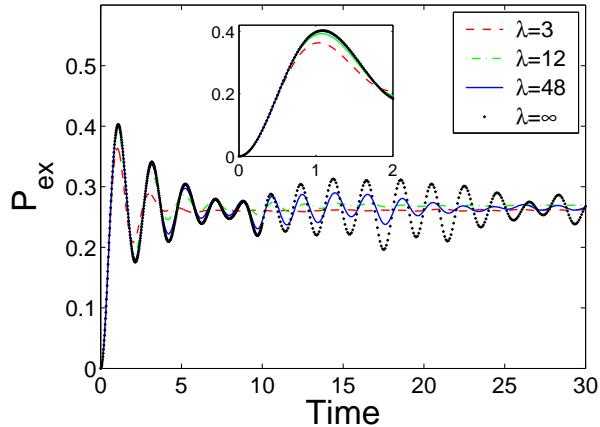}
\caption{ Excitation fraction $P_{ex}$ as function of time for
different $\lambda$. The inset shows enlarged version of initial
dynamics.} \label{Pex_nbar}
\end{figure}

In Fig. \ref{Pex_nbar}, we show the numerical results of
excitation fraction as function of time for pseudoatom number
$N=16$. The results are obtained by using wrap boundary condition
and averaging over 1000 different configurations of $\{J_k\}$. We
can see that the overall trends of the curves for different
$\lambda$ are similar. In all cases, $P_{ex}$ first increases
almost linearly and overshoots to a maximal value, then oscillates
and saturates to about $26\%$, close to the expected value
($25\%$). It comes from the fact that for two pseudoatoms with
perfect blockade, the wave function should behave like $\cos(J
t)|gg\rangle+\sin(J t)(|eg\rangle+|ge\rangle)/\sqrt{2}$, so the
total time-averaged excitation for one pair of the pseudoatoms is
$50\%$, i.e., $25\%$ for each pseudoatom. It's also obvious that
the larger $\lambda$, the less fluctuation in $J_{k}$, so that we
can observe more oscillations. For $\lambda=\infty$, when there is
no fluctuation at all, the oscillation period can be estimated
from analyzing the spectrum of the Hamiltonian. Labelling the
eigenvalues and the corresponding eigenstates as $E_{\alpha}$ and
$|E_{\alpha}\rangle$ ($\alpha=1,2,...,N_b$) respectively, we find
that there exists a nearly periodic structure in the plot of
$|\langle \psi(t=0)|E_{\alpha}\rangle|^2$ versus $E_{\alpha}$. In
this case, $P_{ex}$ does not saturate to a fixed value for $N=16$
pseudoatoms.

The excitation fraction of a physical Rydberg gas can be
qualitatively obtained by mean field approach \cite{gould}, which
is essentially a reduction to single-atom picture. However, the
two-atom correlation, where the off-diagonal terms of the density
matrix are important, cannot be simply accounted for from the mean
field approach. In the following, we will focus on correlation
properties of reduced two-pseudoatom subsystem from a full quantum
mechanical calculation.

\begin{figure}[htb]
\includegraphics[width=3.15in]{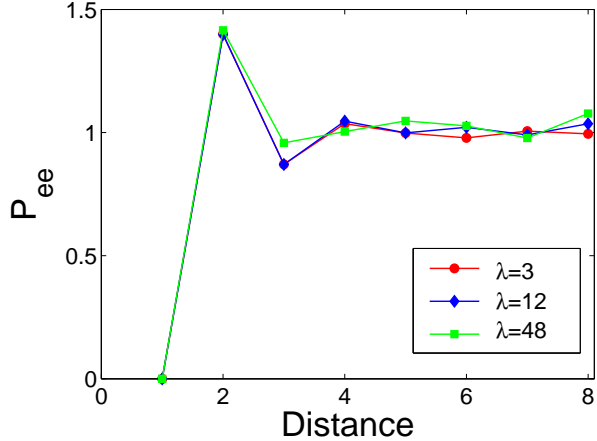}
\caption{ Pair correlation $P_{ee}$ as function of distance for
different $\lambda$. Solid lines are guides to the eye.}
\label{Pee}
\end{figure}

The first quantity we will look into is the pair correlation,
$P_{ee}$, at saturation time, which is defined as the ratio of the
probability of both pseudoatoms excited divided by the square of
probability of single-pseudoatom excitation \cite{Robicheaux}. In
Fig. \ref{Pee}, we show $P_{ee}$ as function of distance for
different $\lambda$. Compared with the results in Ref.
\cite{Robicheaux} (see Fig. 2) which uses a continuous model, our
results captures the main physics in a much simpler way. We can
see that for both the discrete and continuous model, $P_{ee}$
exhibits similar behaviors: it is negligible within a distance and
there is a sharp increase after that; at large distance, it
saturates to 1, corresponding to no correlation. For all three
$\lambda$ in our calculation, the next to nearest neighbor always
has the largest value. This can be interpreted as follows.
Assuming the index of the first pseudoatom to be 0, the separation
$d$ is simply the label of the other pseudoatom. $P_{ee}$ for
$d=1$ is 0 because pseudoatom 1 cannot be excited once pseudoatom
0 is excited. For $d=2$, $P_{ee}$ is well above 1. This is because
if pseudoatom 0 is in the excited state, then pseudoatom 1 must be
in the ground state. Thus pseudoatom 2 is more likely to be in the
excited state. Similarly, pseudoatom 3 will be less likely in the
excited state, resulting in a value lower than 1. The fast
approach to 1 at large distance means that such pair correlation
is only short ranged. We do not present the result for
$\lambda=\infty$ due to the non-negligible oscillation even at
long enough time.

While the above pair correlation only relies on the diagonal
elements of the reduced two-body density matrix, the genuine
quantum correlation must be computed from the full reduced
two-body density matrix. We use entanglement of formation (EOF) as
our measure for quantum correlation \cite{wootters}. It is related
to the so-called concurrence, $C={\rm
max}(0,\sqrt{\lambda_1}-\sqrt{\lambda_2}-\sqrt{\lambda_3}-\sqrt{\lambda_4})$,
where $\lambda_i$ are the eigenvalues of a Hermitian matrix,
$R\equiv \sqrt{\rho} ( \sigma_y \otimes\sigma_y) \rho^*
(\sigma_y\otimes\sigma_y) \sqrt{\rho}$, in descending order. EOF
($\varepsilon$) is then given by $\varepsilon =
h((1+\sqrt{1-C^2})/2)$, where $h(x)=-x \log_2(x)-(1-x)\log_2(1-x)
$. $\varepsilon$ gives 0 for separable states and 1 for maximally
entangled states.

\begin{figure}[htb]
\includegraphics[width=3.15in]{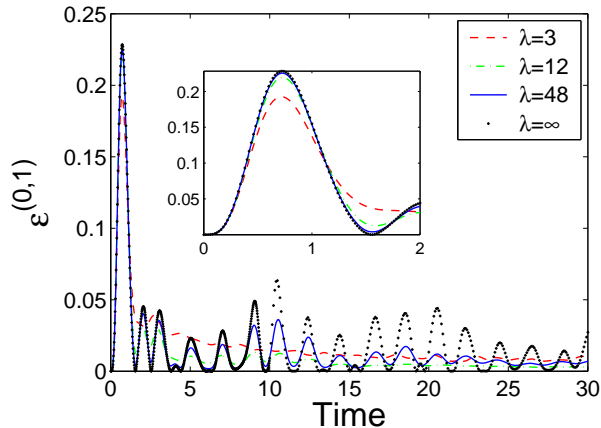}
\caption{EOF $\varepsilon$ as function of time for different
$\lambda$. The inset shows enlarged version of initial dynamics. }
\label{EOF_01_time}
\end{figure}

\begin{figure}[htb]
\includegraphics[width=3.15in]{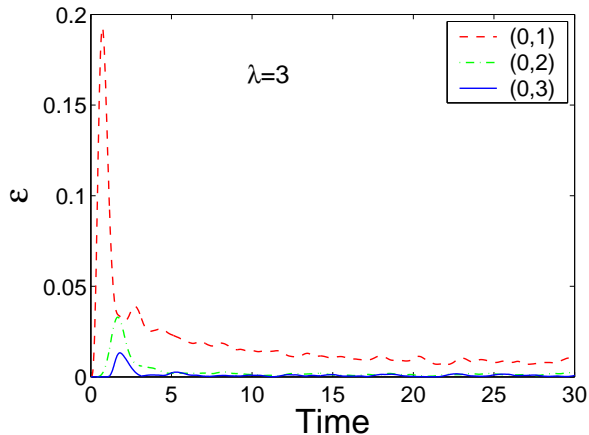}
\caption{$\varepsilon$ as function of time for different pair with
$\lambda=3$.} \label{EOF_pair_time_lambda_3}
\end{figure}

In Fig. \ref{EOF_01_time}, we show EOF for pseudoatom pairs
$(0,1)$ for different $\lambda$. We find that
$\varepsilon^{(0,1)}$ shows similar pattern as excitation
fraction. During the same time scale, they approach a maximal
value and then decay slowly. This reflects two simple facts that
entanglement is originated from excitation and it is superposed
coherently in the initial stage. The behaviors of entanglement for
other pairs are shown in Fig. \ref{EOF_pair_time_lambda_3}. We can
see they again behave in a similar way, although the first peaks
are offset in time.

\begin{figure}[htb]
\includegraphics[width=3.15in]{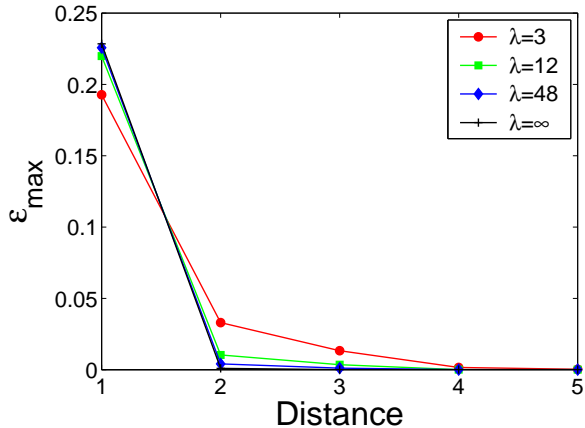}
\caption{The first peak of EOF as function of distance for
different pair. Solid lines are guides to the eye.}
\label{peak_eof_vs_distance}
\end{figure}

To investigate how entanglement varies with the distance between
pseudoatoms, we use the first peak of EOF for different pairs to
characterize its dependence on the distance. Our results are shown
in Fig. \ref{peak_eof_vs_distance}. We note that EOF drops almost
exponentially as the distance increases for all $\lambda$. Thus we
emphasize an important point here: only nearest neighboring
pseudoatoms have relatively large entanglement. Entanglement of
other pseudoatom pairs are negligible during the full time scale
we investigate. This is consistent with our intuitive
understanding that there is no long-range order in such systems.

We can see that quantum correlation is significant only for
nearest neighboring pseudoatoms. However, this is not the case for
the total correlation (including classical correlation). To
compute the total correlation, we adopt the measure as suggested
by D. L. Zhou and L. You \cite{duanlu}. For density matrix
$\rho_{(12)}$ of pseudoatoms 1 and 2, it is given by
\begin{eqnarray}
M_c = {2\over 3}{\rm Tr} |\rho_{(12)}-\rho_{(1)}\otimes
\rho_{(2)}|,
\end{eqnarray}
where $\rho_{(k)}$ is the reduced density matrix of the $k$-th
pseudoatom. Different from Ref. \cite{duanlu}, here the prefactor
$2/3$ is chosen so that this measure gives 1 for a maximally
entangled state.

\begin{figure}[htb]
\includegraphics[width=3.15in]{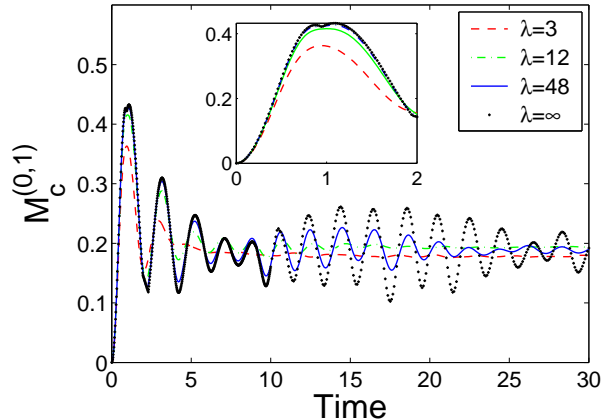}
\caption{Total correlation of pair $(0,1)$ as function of time for
different $\lambda$. The inset shows enlarged version of initial
dynamics.} \label{corr_01_time}
\end{figure}

\begin{figure}[htb]
\includegraphics[width=3.15in]{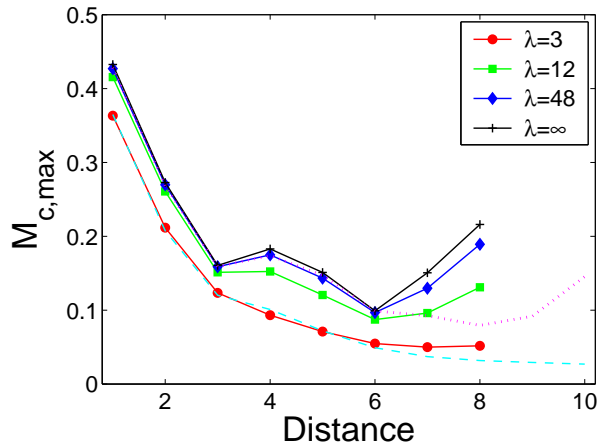}
\caption{The first peak value of total correlation as function of
distance. Marked curves are for $N=16$ with corresponding
$\lambda$ shown in the legend. Unmarked curves are for $N=20$ with
the lower (dashed) and upper (dotted) curve corresponding to
$\lambda=3$ and $\lambda=48$, respectively.}
\label{peak_corr_vs_distance_pair}
\end{figure}

In Fig. \ref{corr_01_time}, we show the total correlation as
function of time for different $\lambda$. We again see that they
show similar pattern as EOF. To investigate the distance
dependence of total correlation, we show the first peak value of
total correlation in Fig. \ref{peak_corr_vs_distance_pair}. We can
see that total correlation does not drop like a single exponential
function with distance as that in EOF. Rather interestingly, they
even do not decrease monotonically as increasing distance. A
simulation with $N=20$ atoms shows similar behaviors with
distance. We find that they give almost identical results for
several initial distances. However, close to the middle point of
the lattice (distance $N/2$), those values of $N=20$ have
correspondingly lower values than $N=16$. Therefore, such
non-monotonicity will become negligible for $N\rightarrow \infty$,
i.e., a finite size effect. We conclude that although quantum
correlation is not important in most cases, the total correlation
cannot be simply ignored, i.e., there is non-negligible classical
correlation in such systems. It is necessary to include not just
nearest neighboring pseudoatoms when considering
correlation-related properties.

The extension of our results to higher dimensions seems difficult
due to the computational limitation. For 2D configurations and
nearest neighboring blockade, we can only simulate up to $5\times
5$ square lattice. Without presenting more figures, we briefly
discuss our results. Our major results of 1D still hold in 2D
configurations. We find that EOF shows similar patterns but drops
even faster as distance increases. This is possibly due to the
fact that in 2D, as more neighboring pseudoatoms are involved, the
fluctuation is more intensive than that in 1D. In addition, the
corresponding pair entanglement is found to be smaller than that
in 1D. Thus we conjecture that the pair entanglement is even
smaller in a 3D cubic lattice.

\begin{figure}[htb]
\includegraphics[width=3.15in]{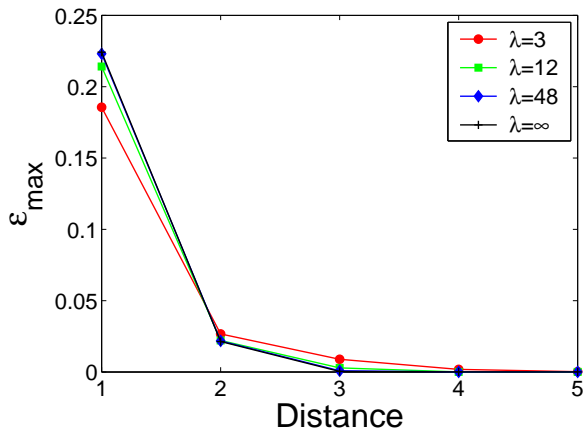}
\caption{$\varepsilon_{\max}$ as function of distance for $N_w=5$
with nearest neighboring blockade.} \label{fulleof}
\end{figure}

\begin{figure}[htb]
\includegraphics[width=3.15in]{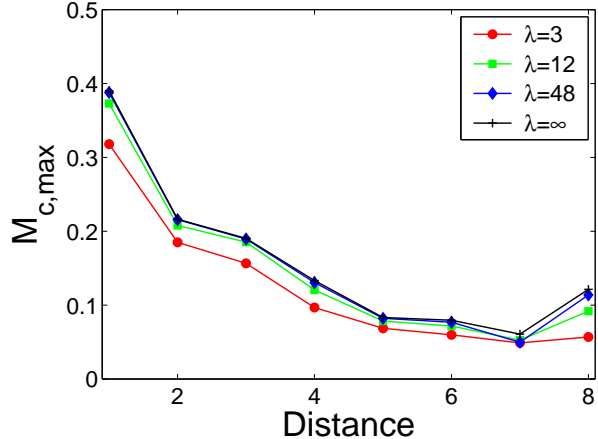}
\caption{$M_{c,\max}$ as function of distance for $N_w=5$ with
nearest neighboring blockade.} \label{fullm}
\end{figure}

Before concluding, we want to justify the validity of our
simplified Hamiltonian. We will show that the fast decay of
entanglement on distance is not an artifact of the assumed short
range interaction. To confirm this point, we carry out simulations
using the long range interaction,
\begin{equation}
\hat{H}'=\sum_k J_k \hat{\sigma}_x^{(k)}+\sum_{j<k}{D\over |j-k|^6
} { 1+\hat{\sigma}_z^{(j)}\over 2} { 1+ \hat{\sigma}_z^{(k)}\over
2},
\end{equation}
where $J_k$ takes the same form as in the Hamiltonian $\hat{H}$.
The dimensionless parameter $D$ quantifies the interaction
strength between two nearest neighboring pseudoatoms. Because we
always use the distance between two nearest neighboring
pseudoatoms as the length scale, $D$ also depends on $N_w$, i.e.,
the decrease in the distance is equivalent to the increase in the
interaction strength. We carry out similar calculations by using
the full basis of $\hat{H}'$. Selected results for $N_w=5$ with
nearest neighboring blockade are shown in Figs. \ref{fulleof} and
\ref{fullm}, for $\varepsilon_{max}$ and $M_{c,max}$ as function
of distance, respectively.  We find that the main features by
using $\hat{H}'$ are qualitatively reproduced by the simplified
Hamiltonian $\hat{H}$, which leads us to conclude that the
behavior of entanglement on distance is not an artifact of short
range interaction in the simplified Hamiltonian $\hat{H}$. And for
the total correlation, the overall shape of the curves remains the
same. More importantly, the total correlation does not drop as
fast as entanglement with distance, which provides another
evidence that our simplified model captures the main physics. We
emphasize that this calculation not only justifies our simplified
model, but also means that those effects in Figs.
\ref{peak_eof_vs_distance}, \ref{peak_corr_vs_distance_pair},
\ref{fulleof}, and \ref{fullm} are mostly due to blockade, which
shows that the long range interaction is actually not so important
to correlation properties.

In conclusion, we have performed full quantum calculations on
correlation properties for pseudoatoms in a 1D lattice structure
with perfect blockade. By comparing the results from a reduced
basis and a full basis calculation, we justify the validity of the
simplified Hamiltonian for perfect blockade among pseudoatoms.
This agreement also means that single- and two-particle
correlation properties (e.g., average excitation, correlation) are
determined mostly by the pure blockade effect. From numerical
simulation, we find that there are both quantum and classical
correlations accompanying the building up of atomic excitation.
Our results show that two-body entanglement is only important for
nearest neighboring pseudoatoms and it drops exponentially fast
with the distance between them even when there is no fluctuation
in the system. However, the total correlation decays much more
slowly with distance, showing the system in this paper is mostly
classically correlated. As a simple extension to higher dimension,
we compute our model system in a two dimensional $5\times 5$
square lattice. We find 2D results agree qualitatively with those
of 1D. From the theoretical point of view, our findings imply that
a better description of Rydberg gas beyond mean field or superatom
picture should also at least take classical correlation into
consideration. We hope that our study can be helpful to the
understanding of Rydberg gases in the strong blockade regime.

We would like to thank J. V. Hern\'{a}ndez for some helpful
discussions. This work is supported by the NSF under grant no.
0355039.

% full author list

\end{document}